# Quantum Critical Points in Ferroelectric Relaxors: Stuffed Tungsten Bronze $K_3Li_2Ta_5O_{15}$ and Lead Pyrochlore ($Pb_2Nb_2O_7$)

Rebecca M. Smith,[1] Jonathan Gardner,[1] Finlay D. Morrison,[1] Stephen E. Rowley,[2,3] Catarina Ferraz,[4] M. A. Carpenter,[5] Jiasheng Chen,[2] Jack Hodkinson,[2] Siân E. Dutton[2] and J. F. Scott[1,6]

1. School of Chemistry, University of St. Andrews, St. Andrews, KY16 9ST, U. K.

2. Cavendish Laboratory, University of Cambridge, J. J. Thomson Avenue, Cambridge, CB3 0HE, United Kingdom

3. Centro Brasileiro de Pesquisas Físicas, Rua Dr Xavier Sigaud 150, Rio de Janeiro, 22290-180, Brazil

4. UFRJ, Estr. de Xerém 27, Xerém, Duque de Caxias, Rio de Janeiro, 25245-390, Brazil

5. Department of Earth Sciences, University of Cambridge, Downing Street, Cambridge CB2 3EQ, U. K.

6. School of Physics, University of St. Andrews, St. Andrews, KY16 9SS, U. K.

## Abstract

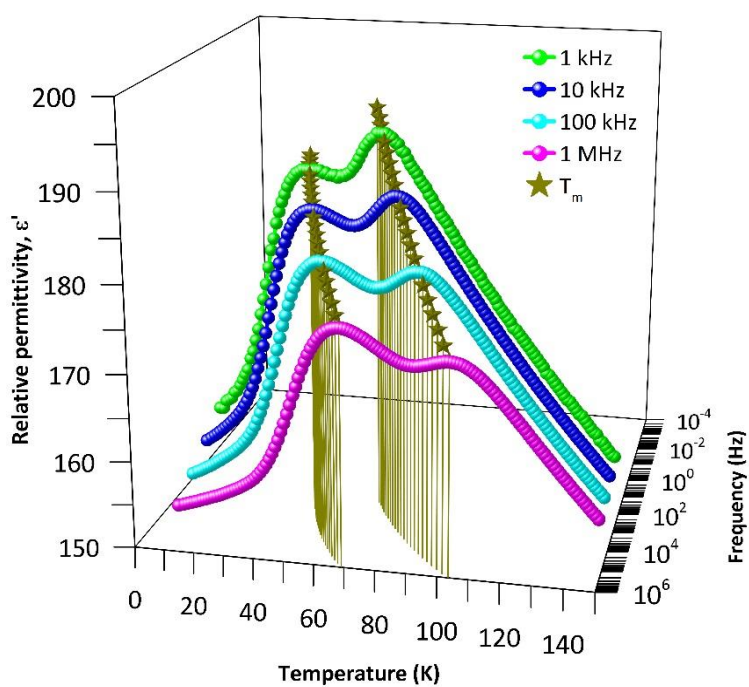

We have synthesised ceramic specimens of the tetragonal tungsten bronze $K_3Li_2Ta_5O_{15}$ (KLT) and characterized its phase transition via X-ray, dielectric permittivity, ultrasonic spectroscopy and heat capacity measurements. The space group of KLT is reported as both *P4/mbm* or *Cmmm* with the orthorhombic distortion occurring when there are higher partial pressures of volatile K and Li used within the closed crucibles for the solid state synthesis. The data show strong relaxor behaviour, with the temperature at which the two dielectric relative permittivity peaks decreasing with 104 K ≥ $T_{m1}$ ≥ 69 K and 69 K ≥ $T_{m2}$ ≥ 46 K as probe frequency f is reduced from 1 MHz to 316 Hz. The data satisfy a Vogel-Fulcher model with an extrapolated freezing temperature for ε' and ε" of $T_{f1}$ = + 15.8 and -11.8 K and $T_{f2}$ = – 5.0 and -15.0 K for f → 0 (tending to dc). Therefore by tuning frequency, the transition could be shifted to absolute zero suggesting KLT has a relaxor-type quantum critical point. In addition, we have reanalysed the conflicting literature for $Pb_2Nb_2O_7$ pyrochlore which suggests that this is also a relaxor-type quantum critical point as the freezing temperature from Vogel-Fulcher fitting is below absolute zero. Since the transition temperature evidenced in the dielectric data at ca. 100 kHz shifts below zero Kelvin for very low frequencies, heat capacity data collected in the zero-frequency (dc) limit, should not indicate a transition. Both of these materials show promise as possible new relaxor-type quantum critical points within non-perovskite based structures as multiple compounds are reported with low-temperature transitions.

I. The search for new quantum critical point (QCP) ferroelectrics including relaxor QCPs

Quantum critical point (QCP) studies of ferroelectrics have been limited to a few crystal structures, emphasizing perovskite oxides. The drive to find new QCPs within ferroelectrics (FEs) are of interest due to the likelihood that they will exhibit novel electrical and thermal properties over wide ranges in temperature and tuning parameters similar to what is seen in more widely studied magnetic counterparts. However as the dynamical exponent is 1 for displacive FE rather than 3 for itinerant ferromagnets, the understanding and modelling of their properties are likely to be more complex, since real systems can exceed the upper and lower critical dimensionality [1]. For a QCP to occur, the transition should be driven by quantum fluctuations rather than classical fluctuations and quantum ones tend to dominate in a region just above 0 K. Interest in ferroelectric quantum critical points has grown rapidly in the past several years, with emphasis upon perovskites [1,2] and several other materials, including hexaferrites [3–6] and organic or molecular crystals [7–10]. However, the QCPs studied thus far do not include many crystal families, and with one recent exception [11], no glassy relaxors. The advantage of relaxor QCPs is that their dielectric permittivity peaks at a temperature that is strongly dependent upon probe frequency (Hz to MHz), and hence the permittivity divergence can be driven exactly through T = 0 via frequency, permitting an extra dimension of experimental phase space to be probed. QCP systems with glassy, highly degenerate ground states at T = 0 are of special interest.

For relaxor ferroelectrics, the Vogel-Fulcher equation [12–14] is used to model the dielectric data. Although this was originally an empirical model extending Arrhenius-type relaxations, it has been retroactively derived theoretically from different assumptions. Two recent and different post facto derivations are given in [15,16]. In the literature such Vogel-Fulcher fits to relaxor data (such as PMN) generally give finite freezing temperatures $T_f$. In the present context, the main point of Vogel-Fulcher modelling to quantum critical point systems is that they give freezing temperatures below absolute zero; this is not unphysical but merely implies that the ground state at T = 0 K is absent of long range order, as expected for QCP.

In the present work we extend such studies to tetragonal tungsten bronze-type structures (TTBs), a broad family of device materials, and pyrochlores. Therefore in a search for new FE QCP, it is preferable to start with FEs with known transitions near 0 K. KLT and $Pb_2Nb_2O_7$ pyrochlore may have QCP transitions with $T_c$ reported as 7 K and 15 K respectively; however, to confirm the behaviour driving the transition, permittivity versus temperature data are required [5,11,17].

II. Tetragonal tungsten bronze-type structures including potassium lithium tantalate

Tetragonal tungsten bronzes (TTBs) have a related structure to perovskites ($ABO_3$) created by rotation of some of the columns of octahedra which maintains a corner sharing network of $BO_6$ octahedra but generates pentagonal A2, square A1 and trigonal C channels along the c axis (most often the polar axes), giving an overall formula of $A2_4A1_2C_4B_{10}O_{30}$ [18]. A-sites are occupied by medium to large cations such as Ba, Ca, K or Na ions, with larger cations preferentially occupying the larger A2-site. The triangular channels are very small and therefore normally empty but can contain Li [19]. Several niobate- and tantalate-based TTBs are of interest, with B ions being Nb or Ta, as they have been reported as ferroelectrics (FEs), both normal and relaxor type. Unfilled TTB of $(Ba_{1-x}(Sr/Pb)_x)_5Nb_{10}O_{30}$ are reported [20,21] as having relaxor behaviour for x<0.5 for Sr/Ba versions with $T_m$ = 333 K at 1 MHz and x = 0.25, whereas x>0.5 and Pb/Ba analogues are normal type FE with $T_c$ = 333 – 823 K. All of the Sr/Ba and Ba/Pb versions have incommensurate structures whether or not relaxor behaviour is seen within the dielectric data. The $BO_6$ octahedra tilts generate the incommensurate superstructures rather than variations in filling sequence with the Ba, Sr or Pb ions and partial occupancies within the A sites. Similarly, empty $Ba_4RE_{0.67}Nb_{10}O_{30}$ (RE = La, Nd, Sm Gd, Dy and Y) TTBs [20] have vacancies reported in some of the A1 and all the C sites. For La, relaxor behaviour is seen with $T_m$ = 297 K at 1 MHz whereas other RE-compounds are reported as normal-type FE with $T_c$ ranging from 406 to 537 K-- a size effect due to decreasing size of RE ions as atomic number increases. The filled $Ba_4Na_2Nb_{10}O_{30}$ TTB has all the A2 and A1 sites occupied by Ba and Na respectively with only the C sites vacant. Within $Ba_4Na_2Nb_{10}O_{30}$ there are a large number of phase transitions, starting at high temperatures with a tetragonal paraelectric phase and finishing at cryogenic temperatures (< 10 K) with a (larger-cell) tetragonal phase. The four or five phases at intermediate temperatures include commensurate and incommensurate (both 1q and 2q modulation) phases and are mostly orthorhombic. These distortions are, however, all within the xy-plane normal to the spontaneous polarisation with 1q modulation being incommensurate in either x or y direction with 2q incommensurate with identical modulation in both x and y directions. So all the phases below the uppermost are ferroelectric (with polarisation P along the z-axis). [Note that after a series of orthorhombic phases at intermediate temperatures, $Ba_2NaNb_5O_{15}$ reverts to a tetragonal *P4nc* commensurate ground state at lowest temperatures [22]; this "reverse symmetry" orthorhombic-tetragonal cooling transition involves a unit-cell doubling.]

$K_3Li_2Nb_5O_{15}$(KLN) is a possible "stuffed" TTB since the stoichiometric material has K on all the A2 and A1 sites and Li on all the C sites; however, several authors [23–27] suggest that an excess of Nb is required to form single phase samples. Therefore the structure either has vacancies and/or mixing of the crystallographic site occupancy. The higher temperature transition around 750 K for KLN is from 4/mmm to 4mm, and the low-temperature transition at 80 K, is from *4mm* to *m*. However in contrast to other previously mentioned TTBs, except Pb containing ones, the lower structural distortion is along the polar axis, not orthogonal to it [28]. Its difference at low temperatures from other tungsten bronzes may arise from its decreased flexibility in the xy-plane due to the stuffed C-channels with Li; therefore the same situation may occur for other "stuffed" TTBs. However, this explanation is not generally true since low-temperature 4mm to m is also reported for TTBs without filled c–channels but with other complicating ions such as Pb or Bi [29–31].

$K_3Li_2Ta_5O_{15}$ (KLT), an analogue of KLN is reported to form as a single phase sample without excess Ta due to the decrease in electronegativity of Ta compared to Nb [23], not due to a size difference, as both ionic radii are 0.64 Å [32]. KLT has been studied as single crystal [33] and dense ceramic [23] samples with the space group of KLT, Figure 1, remains contentious as both *Cmmm* and *P4bm* were reported; however, the latter authors could not determine orthorhombic distortion within their PXRD data. The substitution of Nb for Ta in any of these TTBs has a drastic effect upon its Curie temperature, for example increasing it from $T_c$ = 7 K in KLT to $T_c$ = ca. 753 K in KLN [34–39]. This extremely large shift in Curie temperature for Nb/Ta substitution has not yet been modelled or explained by DFT calculations; note that there is no change in ionic radius, so the large change in $T_c$ is presumably a mass or electronegativity effect. The single crystal's electrical properties showed $T_c$ = 7 K with a polarisation-electric field hysteresis loop occurring below $T_c$ within at 50 Hz. Whether there is a frequency dependence in dielectric permittivity is unknown as both papers studied only one frequency.

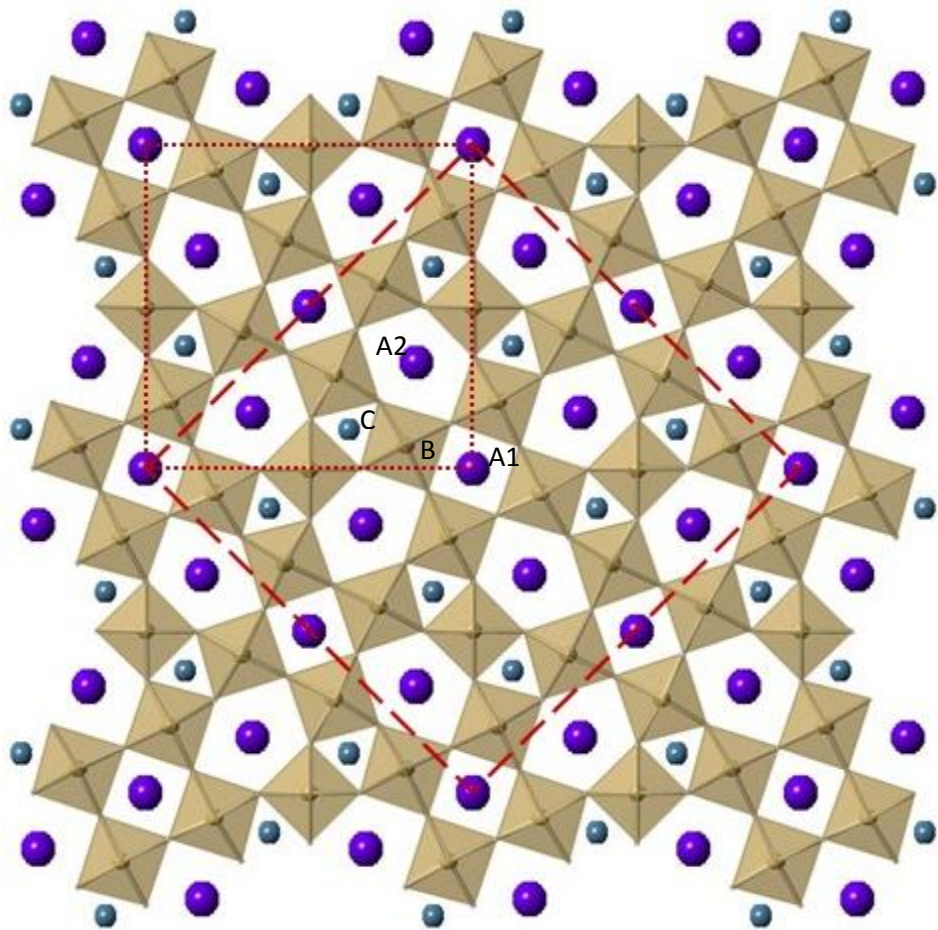

Figure 1: Theoretical [23,33] TTB structures for KLT for ab plane in *P4/mbm* (short dashes) and *Cmmm* (long dashes) space groups with cell parameters for *P4/mbm* as a = 12.60 Å, c = 3.936 Å and *Cmmm* as a= 17.78 Å, b = 17.83 Å, c = 3.931 Å.

Fukuda reports [33] that Ta-rich members of the $K_3Li_2(Ta_xNb_{1-x})_5O_{15}$ solid solution undergo a tetragonal to orthorhombic transition with falling temperatures from 523 K for higher values of x to 295 K for x = 0.55 with the generation of cross-hatch twins above x = 0.7. The symmetry change is most probably *P4/mbm – Cmmm*. It is inevitable that ferroelastic twin walls parallel to (110) and (1-10) of the parent tetragonal structure will be present in orthorhombic KLT crystals which were synthesised at high temperatures and then cooled through the transition point.

### A. Experimental methods

#### *1. Synthesis, PXRD and SEM/EDX*

KLT was synthesised using stoichiometric amounts of dried $K_2CO_3$ (Sigma Aldrich 99%), $Li_2CO_3$ (Alfa Aesar 99%) and $Ta_2O_5$ (Alfa Aesar 99%). The powder was ball milled for 2 h in a planetary ball mill at 600 rpm in ethanol then pressed into loose pellets to be reacted within 4 ml closed crucibles surrounded by sacrificial powder, (composition below) at 1073 K (ramp rate of 10 K min$^{-1}$) for 2 h. Pellets were ground in an agate mortar and pestle before being pressed in a uniaxial press at 500 psi before sintering at 1573 K for 4 h. K and Li are volatile so escape from the powder or pellets during high temperature sintering therefore attempts were carried out with 17% excess or deficient K and Li within the surrounding sacrificial powders compared to stoichiometric KLT therefore creating higher or lower K and Li partial pressures, respectively, within the crucibles. Other than the room temperature PXRD patterns and SEM/EDX analysis, all data is reported on samples made with 17% excess K and Li within the sacrificial powder.

Room temperature PXRD data from pellet surfaces was undertaken on a PANalytical Empyrean diffractometer from 5 – 90° with Cu k$_{α1}$ source and step size of 0.017°. Variable temperature measurements, T = 12 – 300 K, were collected on crushed powder on Bruker D8 Advance diffractometer using an Oxford Cryosyetem PheniX stage from 10 – 110° with step size of 0.0102°using Cu k$_α$ sources, $\lambda$ = 1.54 Å. Rietveld refinements [40] were performed using GSAS/EXPGUI [41,42]. Peak shapes were modelled using a pseudo-Voigt function and the background fit using a twelfth order Chebyschev polynomial. To account for preferred orientation along [00l], a spherical harmonic order 28 model was used to account for changes in intensity.

Scanning electron microscopy (SEM) and energy-dispersive x-ray spectroscopy (EDX) was undertaken on the same polished sample surface as the room temperature PXRD. This analysis used a Jeol JSM 5600 scanning electron microscope with an accelerating voltage of 30 kV and an Oxford Inca EDX system. For EDX analysis, three separate areas of the pellet were used and within each area seven different spectra were collected giving 21 spectra in total. Each spectrum was analysed twice, with and without the O peak being used in the calculations, the values for O as expected seemed to be unrealistic so are not reported. As results were around the stoichiometric KLT ratios, PXRD GSAS/EXPGUI refinements were undertaken without altering K occupancies.

*2. Specific heat data*

The specific heat measurements were undertaken on 2 sets of kit, one of which was on Quantum Design PPMS system using the standard relaxation technique on a pellet sample with a mass of a few mg, the other used crushed powder from a pellet of similar mass.

*3. Dielectric properties and immittance spectroscopy*

The circular faces of pellets for electrical measurements were polished with silicon carbide abrasive paper before applying silver electrode paste (RS components) and curing for ≥ 15 mins at 423 K for permittivity measurements. The dielectric peak maxima were extracted from a 9$^{th}$ order polynomial around each peak and solved iteratively using the Vogel-Fulcher relationship for both relative (real) and imaginary permittivity.

For immittance measurements, gold electrodes were sputtered onto the faces which are able to withstand higher temperatures. Relative permittivity and dielectric loss measurements were performed using an HP 4192A impedance analyser over frequency range from 100 Hz to 1 MHz, from 15 K to 300 K with the sample mounted in a Janis cryostat with closed-cycle helium refrigerator. Immittance measurements using a Wayne Kerr 6500B impedance analyser were undertaken from 303 K to 832 K approximately every 5 K in the frequency range of 25 Hz to 1.28 MHz using custom-made sample holder within a Carbolite furnace. The data were extracted using ZView software to generate an Arrhenius plot and, from linear fits, activation energies determined for each region.

*4. RUS*

The sample used for RUS measurements was in the form of a colourless rectangular parallelepiped with dimensions 2.756x2.757x1.046 mm$^3$ and mass 0.0345 g, which had been cut from a larger ceramic fragment. Rounding of some edges was clearly due to falling away of individual crystals, indicating a slight propensity for disaggregation.

The RUS technique has been described in detail by Migliori and Sarrao [44] and the instrument used in the present study by Schiemer et al. [45] and Evans et al. [46]. Low temperatures are delivered by a cryogen free Oxford Instruments Teslatron cryostat and the RUS head sits in a sample chamber containing a few millibars pressure of helium gas to provide thermal exchange with the sample. The KLT parallelepiped was placed in the RUS head with piezoelectric transducers in light contact with the pair of its largest faces. Primary data were collected in the frequency range 10-1200 kHz with 65,000 data points per spectrum. An automated sequence of cooling and heating, with a settle time of 60 s for thermal equilibration before data collection at each set point, was as follows: cooling from 295 to 5 K in 5 K steps, followed by heating from 4 to 30 K in 2 K steps, 31 to 65 K in 1 K steps, 70 to 295 K in 5 K steps.

Analysis of the primary spectra was conducted offline using the software package Igor (Wavemetrics) to fit individual resonance peaks with an asymmetric Lorentzian function. The square of the peak frequency, $f$, of each peak scales with the elastic constant which determine the distortions involved. The peak width at half maximum height, $\Delta f$, provides a measure of acoustic loss as the inverse mechanical quality factor, $Q^{-1}$, which is taken here to be $\Delta f/f$. The normal modes of the sample are governed predominantly by shearing, typically with only small contributions from breathing motions, and variations of $f^2$ for a polycrystalline sample are therefore dependent predominantly on the shear modulus, with only small contributions from the bulk modulus

B. Results

*1. PXRD and SEM/EDX*

The partial pressures of volatile K and Li affects whether an orthorhombic distortion occurs from *P4/mbm* to *Cmmm* within the room temperature PXRD patterns. SEM/EDX was undertaken on both of these surfaces to determine the ratios for K:Ta with ideal values being 3:5, however, obtained ratios varied depending on whether the observed O peak was removed. .

At lower K and Li partial pressures, the crystal structure refines in *P4/mbm* (wR$_p$ = 0.1132, R$_p$ = 0.0786, $\chi^2$ = 4.642, R(F$^2$) = 0.0535) with no obvious splitting in peaks seen, Figure 2a and attempts at refining in orthorhombic *Cmmm* space group resulted in no major improvements confirming that tetragonal *P4/mbm* is the correct space group based on the current PXRD data. The SEM/EDS ratios are 2.51:5 or 3.01:5 (range 2.31-2.81:5 or 2.83-3.33:5) for with and without O peak used, respectively.

Whereas when higher K and Li partial pressures are used, clear peak-splitting occurs within the PXRD pattern, Figure 2b, which refines in orthorhombic space group of *Cmmm* (wR$_p$ = 0.0976, R$_p$ = 0.0619, $\chi^2$ = 3.012, R(F$^2$) = 0.0470) giving a 0.21% orthorhombic distortion. The SEM/EDS ratios for with and without O peak are 2.60:5 or 3.14:5 (range 2.30-2.89:5 or 2.85-3.41:5) showing a higher K content.

All further results are reported on samples made under higher partial pressures of K and Li because of giving lower PXRD GSAS/EXPGUI refinement parameters and more ideal ratios according to SEM/EDX. Variable temperature PXRD was undertaken from room temperature down to 12 K, which showed no obvious changes in cell parameters but an increase in orthorhombic distortion with decreasing temperature, Figure 2(c).

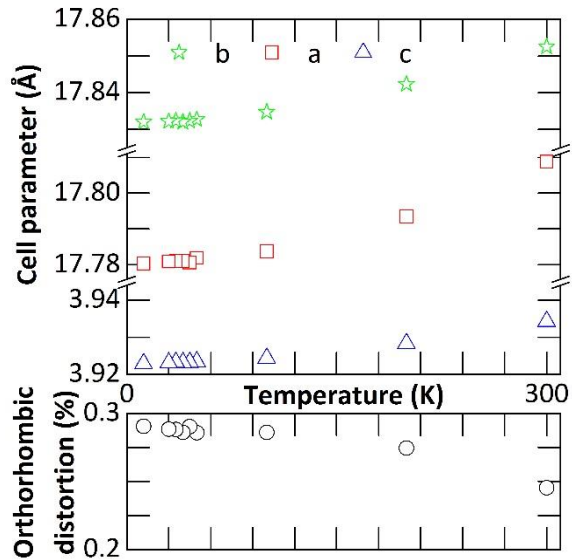

Figure 2: Room temperature PXRD of pellets' surfaces sintered under (a-top) lower and (b-middle) higher partial pressures of K and Li in space group *Cmmm* with an insert (c-bottom) of variable temperature PXRD of powder made by grinding up a higher partial pressure K and Li pellets refined in *Cmmm* space group.

*2. Specific heat capacity*

The specific heat capacity data are shown in Figure 3. The temperature dependence of the heat capacity is cubic at low temperatures before reaching a lower power at higher temperatures. No anomalies were observed over the temperature range measured consistent with the absence of a phase transition in the low frequency (dc) limit.

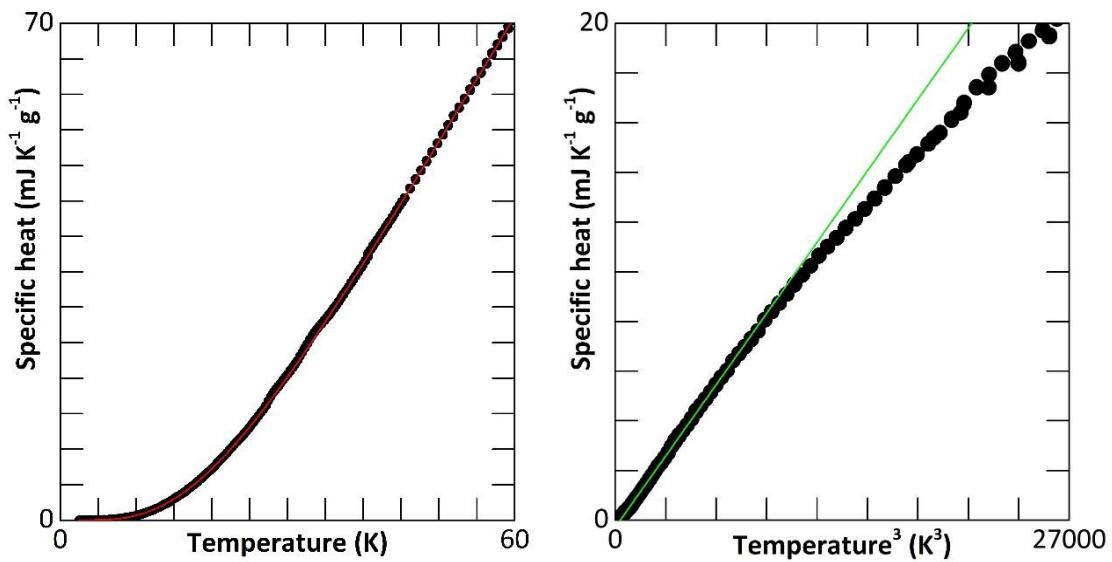

Figure 3: (a) The left figure shows the specific heat capacity plotted versus temperature. No anomaly is observed over the full temperature range measured consistent with a ferroelectric relaxor freezing temperature suppressed below zero. As shown in the right figure a cubic temperature dependence of the heat capacity is observed at low temperatures as expected from the contribution of acoustic phonons.

*3. Dielectric data*

The dielectric data, Figure 4a, showed two frequency dependent peaks in both the relative (real) permittivity, ε', and loss (imaginary permittivity), ε", at temperatures of $T_{m1}$ = 59.9 K in ε' and $T_{m2}$ = 90.9 K at 100 kHz; both peaks exhibit relaxor behaviour. In an aim to characterise this relaxor behaviour several types of fits were undertaken, Equations 1 and Table 1, to model the relaxor-type peak maximums within both the relative and imaginary permittivity data. Vogel-Fulcher fits, Figure 4b, with variable freezing temperatures significantly reduces the error between the experimental maximums and the fitted values (SSR or sum of squared residuals) when compare to Arrhenius fits (where the freezing temperature is fixed to absolute zero). Critical power-law fit is an alternative fit which for ε' does not give as low an error (SSR) or as good a fit by eye as either of the other two types of fit. When extracting the peak maximums, it became obvious that the relative permittivity peaks overlap more than imaginary ones so both were extracted and analysed. The attempt frequencies $f_0$ are quite uncertain. By extrapolation of Vogel-Fulcher fits, the vertical asymptote is the freezing temperature, $T_f$, when an extrapolated glassy structural transition may occur. The horizontal asymptote is the fundamental attempt frequency and is higher value for $T_{m2}$ than $T_{m1}$. The frequency $f_0$ in Vogel-Fulcher fits is not often reliable, because the least-squares fitting is not very sensitive to this number. However, to be physically plausible it should be of order a phonon frequency. Optical phonons of long wavelength are typically $10^{13}$ Hz (300 cm$^{-1}$), and acoustic phonons of order $10^{11}$ Hz (3 cm$^{-1}$), as observed here. The data above the higher temperature dielectric peak fits Curie-Weiss behaviour, Figure 4c, which for the above $T_{m2}$ gives a Curie constant of 70855 - 74147 K depending on frequency however gives an unrealistic Curie temperature of between - 298 and – 319 K.

Vogel-Fulcher: $f = f_0 e^{-\frac{E_a}{k(T_m - T_f)}}$

Arrhenius: $f = f_0 e^{-\frac{E_a}{kT_m}}$

Critical power law: $f = a(T_m - T_0)^b$

Equations 1: Possible fits for relaxor-type behaviour observed in relative and imaginary permittivity used to obtained constants in Table 1.

Table 1: Vogel-Fulcher (VF), Arrhenius (Arr.) and critical power-law (crit.) fits to relative permittivity, ε', from 316 Hz to 1 MHz and a Vogel Fulcher fit for imaginary permittivity from 1 or 1.47 kHz ($T_{m2}$ or $T_{m1}$) to 1 MHz due to limitations of equipment. The errors in calculated values determined from standard deviation error in $T_m$ used.

| Fit type | Parameter (units) | ε' $T_{m1}$ | ε" $T_{m1}$ | ε' $T_{m2}$ | ε" $T_{m2}$ |
|---|---|---|---|---|---|
| VF | $f_0$ (Hz) | 4.54×10$^{10}$ | 2.26×10$^{10}$ | 2.81×10$^{13}$ | 3.97×10$^{13}$ |
| | $E_a$ (eV) | 0.05(1) | 0.08(1) | 0.16(1) | 0.17(1) |
| | $T_f$ (K) | + 15.8(11) | − 11.8(5) | − 5.0(3) | − 15.0(7) |
| | SSR (Hz$^2$) | 0.0267(19) | 0.0041(2) | 0.0062(4) | 0.0060(3) |
| Arr. | $f_0$ (Hz) | 1.41×10$^{13}$ | 4.82×10$^{10}$ | 8.53×10$^{12}$ | 1.46×10$^{13}$ |
| | $E_a$ (eV) | 0.10(1) | 0.05(1) | 0.14(1) | 0.12(1) |
| | SSR (Hz$^2$) | 0.2485(135) | 0.0653(30) | 0.0121(7) | 0.0098(5) |
| Crit. | $T_0$ (K) | 42.02(228) | | 60.04(326) | |
| | a | 1.06×10$^{-2}$(6) | | 2.07×10$^{-4}$(11) | |
| | b | 5.56(30) | | 5.89(32) | |
| | SSR (Hz$^2$) | 1.18×10$^8$(6) | | 6.45×10$^9$(35) | |

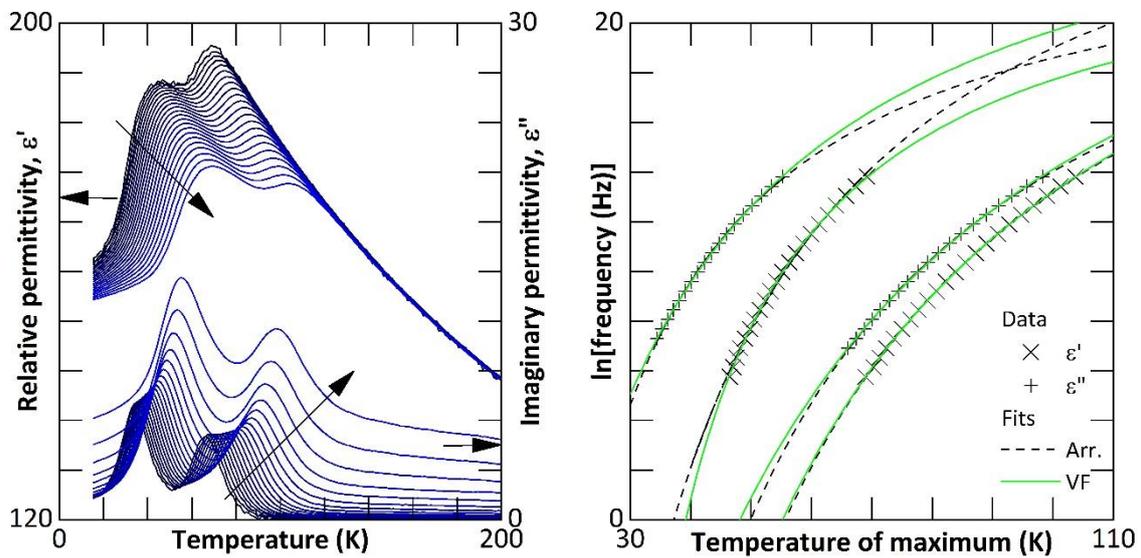

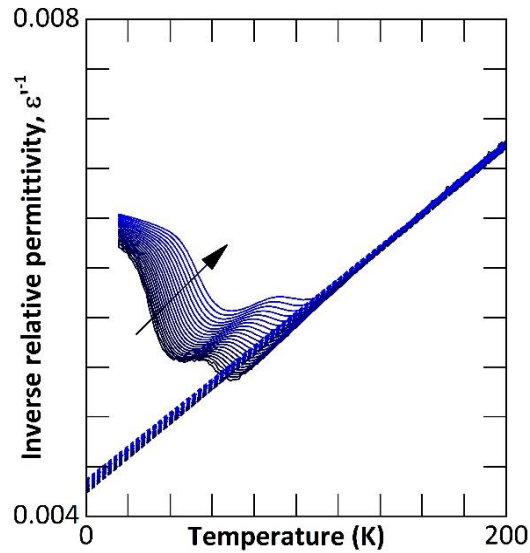

Figure 4: (a-top left) relative (real) permittivity, ε', and imaginary permittivity, ε", on heating with arrow showing increasing frequency (316 Hz - 1 MHz) (b-top right) Vogel Fulcher fits to both relative and imaginary permittivity peaks giving values reported in Table 1. (c-bottom) Curie-Weiss behaviour plot above the higher temperature dielectric peak with fitting of linear to obtain Curie constant of 70265 - 74129 K depending on frequency fitted to range above $T_{m2}$ however gives an unrealistic Curie temperature of between - 319 and – 294 K.

*4. Immittance spectroscopy*

Immittance spectroscopy generated semicircles and peaks within measurement range above 575 K, Figure 5a. From the Arrhenius plot, Figure 5b, the activation energies were determined for the two different regions of 575 – 671 K and 723 – 840 K as there is a change in gradient between these regions. The activation energy for the region of 575 – 671 K is 0.90, 0.90, 0.89 and 0.89 eV for M", Z", M* and Z* formalisms, respectively. Whereas the activation for the higher temperature region is 0.770, 0.757, 0.785 and 0.762 eV similar for M", Z", M* and Z* formalisms.

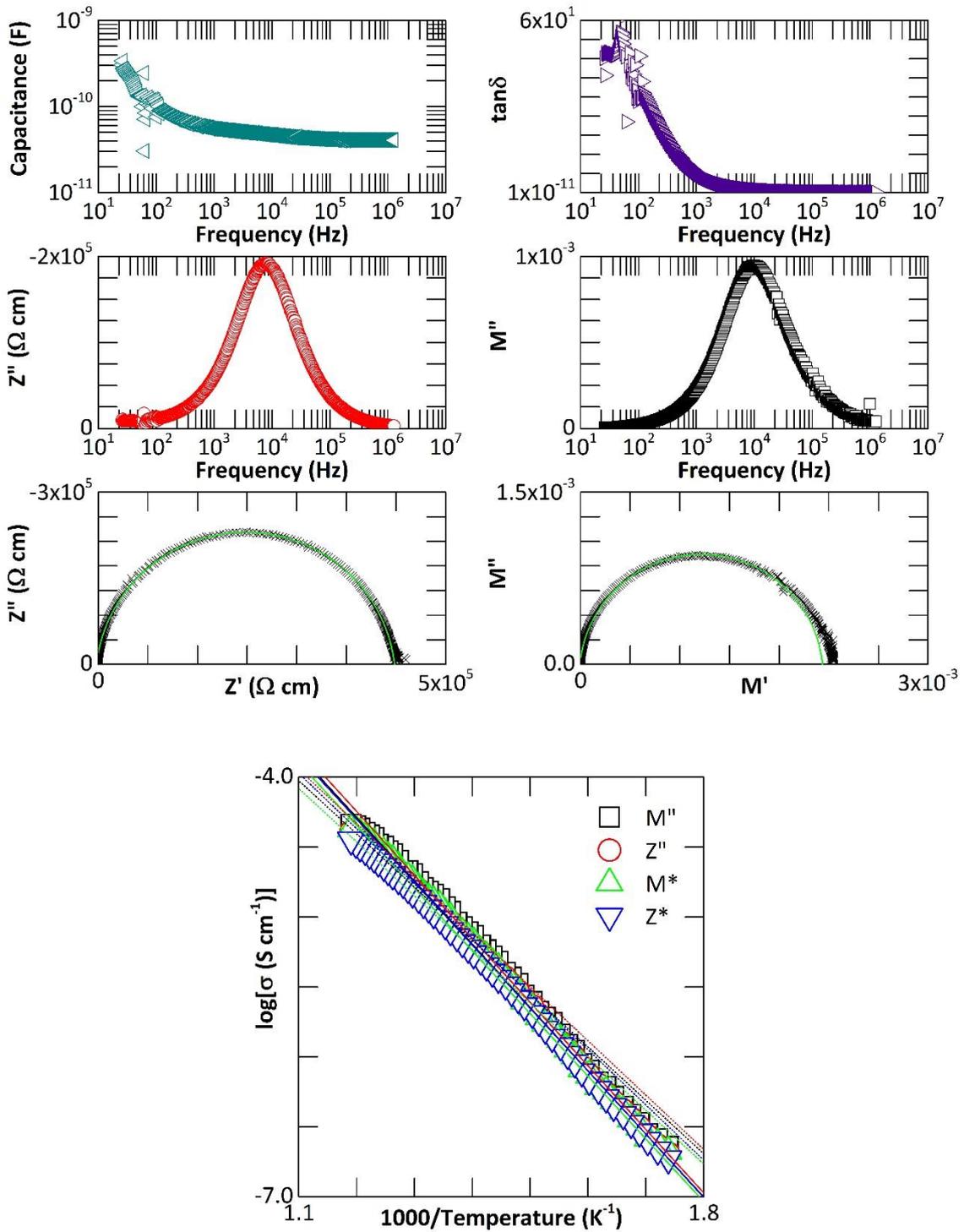

Figure 5: Imittance spectroscopy plots (a-top) example of Nuquist and Bode plots at 671 K with single semicircles corresponding to domain permittivity and a relaxation time of t ≈ 18 μs; R ≈ 0.4 MΩ; and C ≈ 47 pF. We interpret this as a response within a grain and no evidence of grain boundary relaxation. The data is extracted to generate an Arrhenius plot (b-bottom) from which activation energies can be determined.

## 5. RUS

Segments of the primary RUS spectra are shown as a stack in Figure 6. Each spectrum has been offset up the y-axis in proportion to the temperature at which it was collected during cooling, and blue curves are fits of the asymmetric Lorentzian function used to determine values of $f$ and $\Delta f$. There is a tendency for all the resonances to appear in pairs because of a near degeneracy arising from the close similarity in two of the edge dimensions of the sample. All the resonances follow essentially the same trend of significant elastic softening with falling temperature to a clear minimum at ~50 K, followed by recovery.

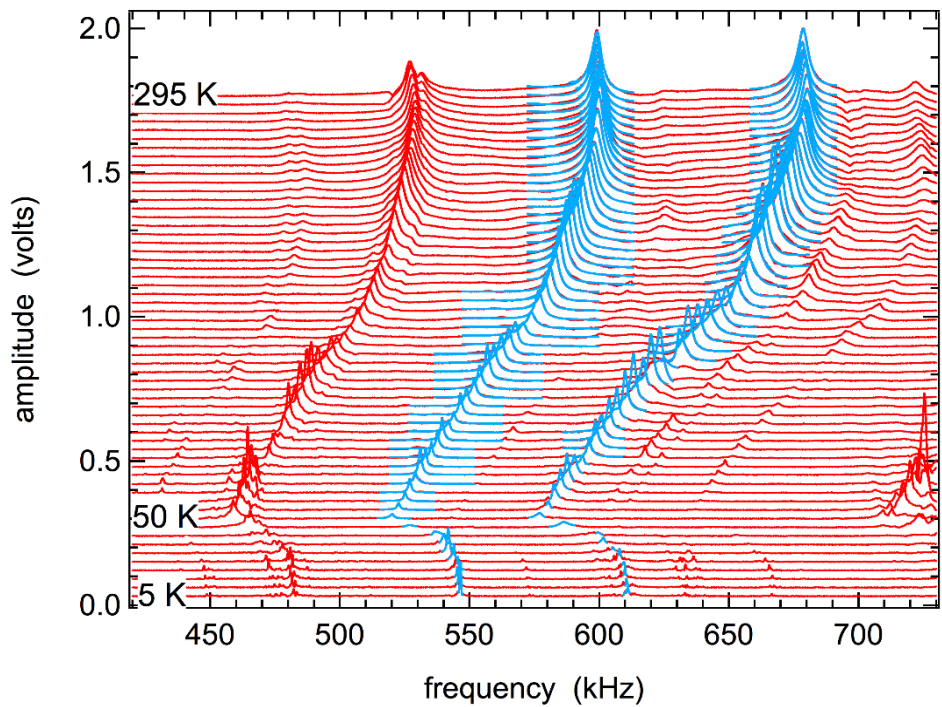

Figure 6: Stack of primary spectra collected at 5 K intervals during a cooling sequence. The left axis is the amplified signal from the detecting transducer and the bottom axis is the frequency of the ac signal applied to the driving transducer. Each spectrum has been been offset up the y-axis in proportion to the temperature at which it was collected. Blue curves are fits of the asymmetric Lorentzian function used to determine values of $f$ and $\Delta f$.

Figure 7a shows variations of $f^2$ and $Q^{-1}$ throughout the complete sequence of cooling and heating for resonances with frequencies near 65, 550 and 600 kHz. There are four distinctive features. Firstly, there is an initial trend of reducing $f^2$, corresponding to softening of the shear modulus, as temperature is lowered, with a maximum softening of ~30%. This ends with an abrupt change to a trend of stiffening by up to ~15% below ~50 K. Slightly regular variations in $f^2$ through ~50 K are accompanied by a peak in $Q^{-1}$ which returns to baseline values below ~25 K and above ~65 K. Secondly, there is difference of ~5 K between heating and cooling for the temperature at which the maximum in $Q^{-1}$ occurs (~40 K cooling, 47 K heating in the case of the peak near 65 kHz) and for the temperature at which the main break in slope of $f^2$ occurs (45 K cooling, 52 K heating, ~65 kHz). Thirdly, there both $f^2$ and $Q^{-1}$ display a distinct hysteresis in the temperature interval between ~50 K and room temperature. The elastic stiffness is lower and the loss higher during heating than during cooling. Finally, there is a gradual increase in the baseline values of $Q^{-1}$ with increasing temperature and there is a broad peak near 250 K during cooling and near 280 K during heating.

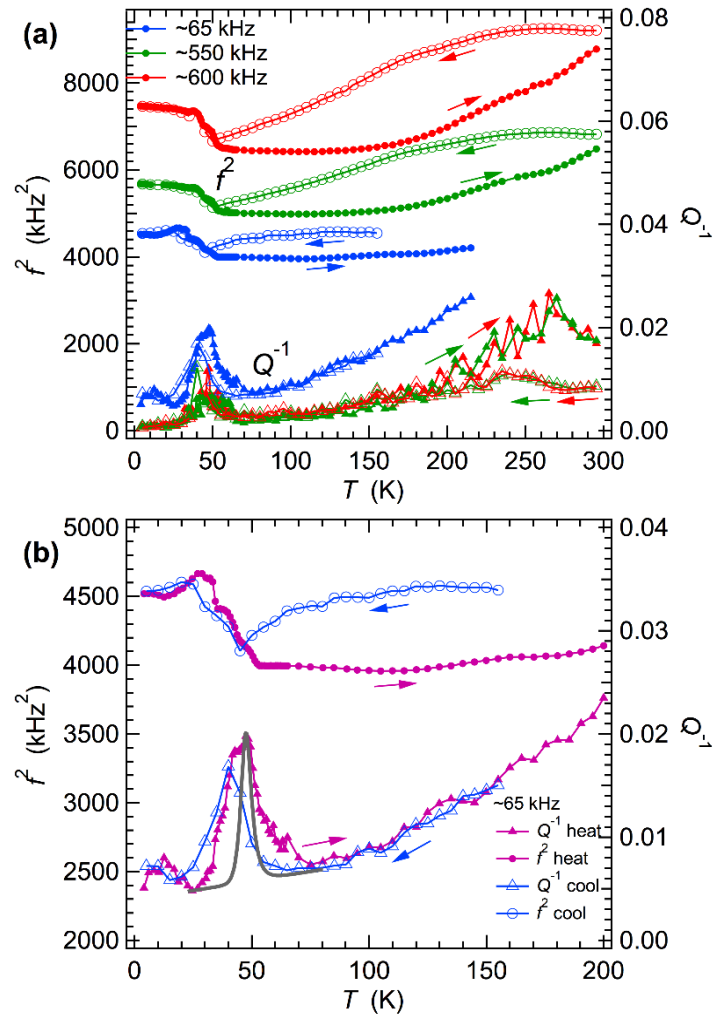

Figure 7: (a) Variations of $f^2$ (circles) and $Q^{-1}$ (triangles) from three different resonances during cooling (open symbols) and heating (filled symbols); $f^2$ values have been rescaled so that data for all three resonances can be compared on the same diagram. (b) Details of the resonance with frequency near 65 kHz. Arrows indicate the direction of changing temperature in the cooling and heating sequences.

Details of the changes in $f^2$ and $Q^{-1}$ are shown for the resonance peak near 65 kHz in Figure 7b. The irregular pattern of variations in the interval ~25-65 K appears to show a sequence of step-like changes in $f^2$ accompanied by a series of overlapping peaks in $Q^{-1}$, which would correspond to a sequence of, perhaps, three Debye freezing events. These can be represented, for the variation of temperature at approximately constant frequency, by [47–50]:

$$Q^{-1}(T) = Q^{-1}_{max} \left[ \cosh \left[ \frac{E_a}{R r_2(\beta)} \left( \frac{1}{T} - \frac{1}{T_{max}} \right) \right] \right]^{-1}$$

.

$Q^{-1}_{max}$ is the maximum value of $Q^{-1}$, occurring at $T_{max}$, $E_a$ is an activation energy and $r_2(\beta)$ is a width parameter to describe a spread of relaxation times (as set out in Table 4-2 of Nowick and Berry [51]). The value of $r_2(\beta)$ is 1 for a single relaxation time but increases as the range of relaxation times broadens. A single smooth curve shown in grey in Figure 7b has $Q^{-1}_{max}$ = 0.0145, $T_{max}$ = 47.5 K and $E_a/r_2(\beta)$ = 0.1 eV. The overall peak in $Q^{-1}$ could be represented by three or four overlapping peaks with about the same activation energy or by a single loss process with $E_a/r_2(\beta)$ = 0.027 eV. If $E_a$ is taken to be 0.1 eV, by comparison with results from dielectric spectroscopy results, $r_2(\beta)$ = 3.71 is obtained. The correlation of step-like increases at ~35, 42 and 48 K in $f^2$ with the location of possible overlapping peaks in $Q^{-1}$ suggests that the correct description would be of a sequence of Debye-like freezing processes.

## C. Discussion

Within the existing prior literature, both space groups *P*4/*mbm* and *Cmmm* are reported [23,33] for ceramic and single crystal KLT samples using XRD analysis at room temperature. The PXRD room temperature results reported here suggest that both space groups can exist depending on the reaction conditions used with KLT syntheses with higher partial pressures of volatile K and Li showing an orthorhombic distortion. Using SEM/EDX, it is possible to determine the K:Ta ratios with and without accounting for the visible O peak, but not Li, within pellets made under these conditions. This shows that the higher partial pressures results in higher amounts of K within the polished pellet's surface when compared to pellets sintered with lower partial pressures implying that the orthorhombic distortion maybe linked to K content within the structure. Orthorhombic distortions can occur when the A1 site distorts from the perfect square channel, and in this case with K on the A1 site would explain the existence of orthorhombic distortions for higher K content. Within TTBs, it is not unprecedented to have orthorhombic distortion and expansions of the unit cells from *P*4/*mbm* (Z = 2) cell; other examples [52–58] include $Ba_2NaNb_5O_{15}$ or $Ba_4La_{0.67}Nb_{10}O_{30}$. $Ba_2NaNb_5O_{15}$ has five structural phase transitions as the temperature is lowered, some incommensurate, all of which are orthorhombic distortions in the ab-plane. From the analysis reported here, there is a high likelihood that at least one phase transition occurs in the region of 25 – 50 K, variable temperature PXRD attempted to probe any possible changes in structures. This suggests no systematic change in cell parameters or refinement parameters exist for KLT. However, this does not exclude that a phase transition occurs within this range, as subtle octahedra tilts rely on exact refinement of O positions such as reported [59] in $Ba_4(La_{1-x}Nd_x)_{0.67}Nb_{10}O_{30}$ are not likely to be seen in PXRD analysis especially in KLT due to the presence of heavier Ta. In addition, these subtle tilts effect the ferroelectric properties as shown [60] in $(Sr_xBa_{1-x})_5Nb_{10}O_{30}$. Neutron diffraction studies are required to probe O and Li effect on structure further.

It is important to try to reconcile the specific heat data and the dielectric data. Specific heat is a quasi-static, not dynamic, measurement for probing any phase transition. This is equivalent to probing the sample in the zero-frequency (dc) limit; therefore the phase transition, in theory, would occur at the freezing temperature $T_f$ extrapolated from dielectric data. However, a feature for $T_{f1}$ is not visible within the measurement window as the linear temperature dependence follows a polynomial trend line. When the specific heat is plotted against temperature cubed, no features are seen except that the Debye temperature, where classical fluctuations dominate and data would plateau, is well above the freezing temperatures from Vogel-Fulcher fits suggesting quantum fluctuations are involved within the dielectric transitions seen which is suggestive of QCPs.

The transitions reported here are above those in the literature. For the single crystal data, [33] $T_c$ is reported as 7 K at an unknown frequency with samples being stoichiometric for both KLT and KLN. This claim for KLN is in direct contradiction to later papers as Fukuda reports [33,61] that KLT and KLN have completely filled C sites with Li and both A1 and A2 with K which is a direct contraction to later papers which say an excess of Nb is required for KLN to form [23–27]. Therefore, there is a slim chance that the lower reported transition could be due to contamination, as widely known for $BaTiO_3$ or deviation from stoichiometric KLT within samples, however without further information resolving this difference is not possible. For ceramic samples, [23] similar data down to 100 K are seen especially within samples that have *P*4/*mbm* symmetry at lower partial pressures. New to this work is the two frequency dependent peaks suggesting relaxor-type dielectric properties for KLT, which is very similar to KLN. As after a careful literature search, KLN's ferroelectric transition is already known to be frequency dependent, with dielectric permittivity maximum shifting around 750 K as reported by Ivett [39]. These two relaxor-type peaks within KLT may be explained by partial occupancies or mixing of the sites within the TTB as both explanations reported as occurring within KLN [27,62], alternatively incommensurate transitions may occur as reported [20] within other TTB compounds. However, which of these exist in KLT will be the subject further investigations and possible subsequent paper.

The Vogel-Fulcher relationship is the best model attempted for the experimental data. The Vogel-Fulcher fits suggest two different relaxation mechanisms occur within KLT with the activation energies and attempt frequencies being of a similar order within the relative and imaginary permittivity for each mechanism. The large variation in the freezing temperatures between ε' and ε" is a reflection that the ε' $T_m$s occur at a higher temperature than ε" $T_m$s shifting the whole Vogel-Fulcher fit to higher temperatures in ε' without altering the shape particularly. The most likely explanation is that ε' peaks were less well defined and separated than in imaginary permittivity which is likely to create higher errors in ε' $T_m$s as in theory the peaks should occur at the same temperature. The freezing temperatures are of interest as a prediction of where the structural transition would occur, if it existed. $T_{f1}$ is determined as +15.8 and – 11.8 K for ε' and ε", respectively, therefore further analysis is required to determine if this structural transition is observable whereas $T_{f2}$ is negative as either – 5.0 or – 15.0 K (ε' or ε") therefore structural transition is unlikely to occur. Negative freezing temperature has been seen previously within hexaferrites [11] and explained by a degenerate glassy ground state occurring at absolute zero generated by the existence of quantum fluctuations. Therefore probing this sample with very low frequency would, in theory, shift the transition to absolute zero suggesting that there is at least one relaxor-type QCP within KLT. $T_{f1}$ for $\varepsilon'$ was within the range of VT-PXRD; however, for reasons described earlier, it is not detected. The Curie constant is calculated as ≈72000 K which is less than the 104800 K reported [63] for $BaTiO_3$ but within the range of other reported oxide ferroelectrics [64].

Within the immittance spectroscopy, the responses are interpreted as within the grains with a small electrode spike at low frequency and higher temperatures, no evidence of grain boundary relaxation is observed. The Arrhenius plot generates two activation energies which is similar to KLN reported by Jun et al. [65]. The lower temperature region, approx. 600 – 700 K, as reported activation energies of 0.90 and 0.83 eV for KLT and KLN respectively. However the activation energy for KLT of 0.67 eV in the higher 723 – 840 K region is comparable to 0.66 eV for KLN reported by Kim et al. [66] but in the lower region of 475 – 600 K suggesting similar mechanisms exist but occur within frequency window measured at different temperatures. As suggested for KLN, the two mechanisms seen in KLT are likely to be linked by Li ion motion even though the activation energies reported here are less than the 1.27 eV for Li ionic conduction reported [67] in $LiTaO_3$ but comparable for other TTBs [68] containing Li with activation energies of around 0.80 eV. The activation energies for KLT at >0.6 eV is larger than the reported [69,70] 0.15 and 0.12 eV for $LiO_2$ and $LiCO_3$ however, for commercial applications, the ionic current per unit weight is a key parameter, and tantalate or niobate tungsten bronzes are not competitive with lighter weight Li oxides, carbonates, or sulfides.

If there is a phase transition in KLT at low temperatures, it does not conform to the normal expectations of strain coupling. A ferroelastic transition with linear/quadratic strain/order ($\lambda eQ^2$) would be expected to give significant elastic softening below the transition point [71,72], but this is not observed. Bilinear coupling of the driving order parameter with the symmetry breaking shear strain for a discrete phase transition ($\lambda eQ$) would give rise to the standard pattern of softening to a minimum at the transition point followed by recovery below it, but, as discussed above, the only direct evidence for a ferroelastic transition is that it would occur above room temperature. The initial trend of elastic softening with falling temperature seen in Figure 7 is more likely to have a dynamic origin and the best analogy may be with the softening seen in RUS data from $Pb(Mg_{1/3}Nb_{2/3})O_3$ (PMN) [73]. In PMN the softening is attributed to the development of dynamical polar nano regions (PNR's) between the Burns temperature of ~630 K and the temperature interval within which they freeze, 230-270 K. The relaxor freezing process is then accompanied by elastic stiffening and a peak in the acoustic loss. However, the latter extends down to at least ~10 K, rather than occurring more discretely in the much narrower temperature interval of ~40 K seen here for KLT.

The pattern for elastic and anelastic effects in KLT thus appears to be of softening due to dynamical effects, such as the condensation of PNR's, followed by discrete pinning or freezing of defects which couple with strain. Changes in dielectric constant and dielectric loss in PMN have the same general form as changes in elastic compliance and acoustic loss but with significant differences in detail, and this was attributed to the fact that the ac electric field produces responses primarily from 180° twin walls, which are not coupled with shear strain, whereas an ac stress produces responses only from 90° twin walls [73]. In KLT, there are two dielectric anomalies, centered at 46 and 69 K when measured at 316 kHz, and one elastic anomaly, centered at ~47 K measured at ~65 kHz. The simplest explanation of the discrepancy is that the higher temperature anomaly involves only 180° switching of local ferroelectric moments while the lower temperature anomaly involves 90° switching, as if the two types of boundaries between PNR's or two types of ferroelectric domain walls freeze at different temperatures. Activation energies extracted from the RUS data are sufficiently close to those extracted from the dielectric data to suggest that the same pinning or freezing process is being sampled.

There are also significant differences between RUS results from KLT and PMN, such as the hysteresis seen for KLT. It is possible that the differences in $f^2$ and $Q^{-1}$ between cooling and heating in the temperature range ~50-295 K are due to opening up of grain boundaries associated with changes in volume due to structural changes below ~50 K. This effect has been seen, for example, when a polycrystalline sample of quartz with grain sizes in the range 0.1-0.3 mm is heated through the α – β transition [74], except that opening of the grain boundaries causes the polycrystalline sample to become elastically softer rather than stiffer, as here. The broad peaks at ~270 K in the heating sequence and at ~240 K in the cooling sequence (Figure 7a), are reminiscent of broad peaks seen at ~200 K in RUS data collected from $Ba_2NaNb_5O_{15}$ (BNN) [75] and an alternative explanation is that the loss is associated with pinning (during cooling) and unpinning (during heating) of ferroelastic twin walls in both materials. This would be analogous to the anelastic behaviour of twin walls in improper ferroelastic perovskites [76]. The sequences of structural changes in BNN and KLT are not the same but a feature in common is that both could undergo tetragonal – orthorhombic transitions during cooling such that both would contain ferroelastic twin walls. At room temperature the twin walls form a classic tweed pattern in BNN [77,78] while they appear for x ≳ 0.7 but are visibly more sparse in $K_3Li_2Ta_5O_{15}$ [33]. There appear to have been no previous optical studies of KLT, but examination of crushed grains mounted in oil using a polarised light microscope has provided some indication that the expected ferroelastic twinning was present in the sample of KLT used for the present study. Changes in the proportions of differently oriented ferroelastic twin domains in a single crystal would also cause changes in the effective elastic constants of the crystal as a whole. Hysteretic effects would then arise in KLT if reorientation of twin walls occurs within individual grains of a ceramic sample, rather than opening up of grain boundaries, in response to stresses at grain boundaries which develop as a consequence of the structural changes at ~50 K and below.

To conclude, the tetragonal tungsten bronze-type structure of KLT has a room temperature structure that depends on the synthesis conditions. With lower partial pressures of K and Li given lower K content, a tetragonal *P4/mbm* structure is observed whereas at higher pressures given near stoichiometric values for K:Ta orthorhombic *Cmmm* structure is obtained. Characterisation of the higher partial pressure near stoichiometric pellets showed that properties depend on both method of analysis and frequency used occurring near absolute zero therefore not always detected. Two dielectric peaks are seen within the range 316 Hz to 1 MHz which fit Vogel-Fulcher analysis with extrapolation give transition temperatures above and below absolute zero. The above 0 K transition may be linked to the feature in specific heat and/or RUS data and a change in structure. However, further work is required to reconcile these differences. The transition that is extrapolated to below 0 K, if it existed, is explained by a degenerate glassy ground state at 0 K created by quantum fluctuations which suggests a relaxor-type ferroelectric QCP occurs in KLT. Thus studies of quantum critical points should be extended to such tungsten bronzes, pyrochlores, and other perovskites.

III. Pyrochlore Lead Niobate

A. Pyrochlore structure

Similar results are seen within pyrochlores $Pb_2Nb_2O_7$ and $Cd_2Nb_2O_7$. The structure of $Cd_2Nb_2O_7$ is a cubic *Fd*-3*m* pyrochlore however due to the increases size and/or lone pair on Pb, a stacking fault occurs every 9 layers creating a trigonal *P*3*m*1 pyrochlore structure [79]. Within $Pb_2Nb_2O_7$ a long-term controversy has existed with Hulm [80] and, independently, Pepinsky and Shirane [81] discovering in lead pyrochlore a ferroic (probably not ferroelectric as "quite" linear hysteresis loop) transition at T ≈ +15.4 K but Siegwarth et al. [82] arguing vehemently that there is no such phase transition, arguing that these are electret effects. It is now generally established that cadmium niobium pyrochlore is glassy below 18 or 19K, but the lead pyrochlore is more enigmatic and controversial. We reconcile these claims by reanalysing the literature to show that this material is also a relaxor QCP, with a dielectric peak at T ≈ 15.4 K at kHz probe frequencies f, but no anomalies for T > 0 in specific heat or XRD measurements, which are essentially dc (f = 0). Hence this pyrochlore is also a relaxor QCP that can be frequency tuned to exhibit dielectric permittivity maximum at T = 0 K.

B. Pyrochlore reanalysis literature section

Similar to the situation in KLT, there was for some years a controversy regarding pyrochlore $Pb_2Nb_2O_7$, with Hulm, Pepinsky and others reporting [80,81,83,84] a phase transition (ferroelectric) at T = 14.0-15.4 K but Siegwarth et al. [82] and Lawless [85] arguing strongly that this was some sort of defect relaxation and not a phase transition at all and hence that Cd-niobate pyrochlore is ferroelectric but Pb-niobate is not.

Our interpretation of the literature is different to Siegwarth et al. [82], shown in Figure 8 and 9: This is not relaxation of defects, but a relaxor ferroelectric transition; our Vogel-Fulcher fit to their original data is displayed in Figure 8. Note that this very low-temperature transition is different from the relaxor transition [86] in $Cd_2Nb_2O_7$ at $T_c$ = 196 K.

Ubic and Reaney show [79] that oxygen vacancies in typical lead pyrochlores result in a lowering of symmetry to trigonal. Thus data can be sample-dependent. This also suggests that lead-niobate pyrochlore may be trigonal below 15.4K. Jayaraman et al. [87] showed in a rarely cited study on an excellent single crystal of stoichiometric $Pb_2Nb_2O_7$ pyrochlore that the ferroelectric phase transition could be reached at a hydrostatic pressure of 4.5 GPa and that after the release of such applied pressure the structure reverted to rhombohedral (trigonal). We speculate that the phase above 4.5 GPa at 292K is the same ferroelectric phase as that below T = 15.4K at ambient pressure. Shrout and Swartz show [88] that lead magnesium niobate also has a relaxor-like transition at a similar cryogenic temperature (T = 20K). The phase diagram in $Cd_2Nb_2O_7$ is already known to be complicated [89–98], with seven phases, including an incommensurate one that locks in to monoclinic below T(lock-in) = 46K, and a glassy phase below $T_g$ = 19K; but $Pb_2Nb_2O_7$ has been more controversial. It is already known that mixed pyrochlores of formula $(Cd,Pb)_2(Nb,Ta)_2O_7$ can be quantum ferroelectrics with Curie temperature $T_c$ passing through zero. [99] We note that the lanthanide pyrochlores also have phase transitions [100,101] in the temperature range near T = 20K; these have been interpreted as magnetically driven, but we suspect that they are typical of most pyrochlores and independent of magnetism.

Siegwarth and Lawless report a very small pyrochlore polarization of 6 nC cm$^{-2}$ for Pb-niobate. He published details of his equipment, in 1971, with such sensitive capabilities. [102] However, although his equipment was very good, his samples were not: In a new 2017 book, [103] it is pointed out that Siegwarth and Lawless never achieved a densification greater than 75% for their lead pyrochlore samples. And perhaps even more important, Sekiya et al. [104] find, unlike most authors, that $Pb_2Nb_2O_7$ is ferroelectric; they conclude in their article that the rapid quenching is important (metastable phases) and that the samples also require a high-temperature anneal after growth, following which their dielectric constant rapidly increases. In addition, the annealing temperature will affect whether a mixed or single phase formed [105]. So the conclusion is that ferroelectricity in lead pyrochlore may depend upon stoichiometry, densification, and annealing procedure.

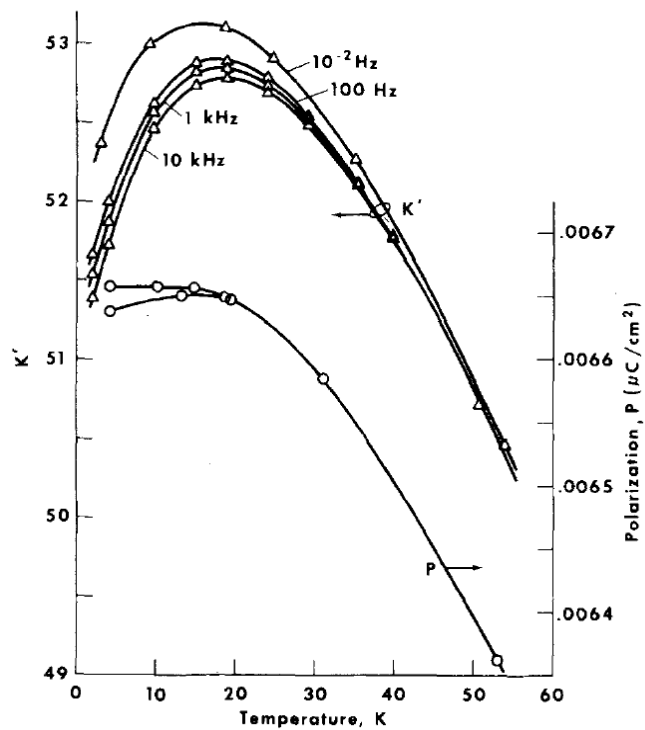

Figure 8: Data of Siegwarth et al. [82] Note the extremely small polarization of nC/cm². [86] gives 2.1 μC/cm² for $Cd_2Nb_2O_7$ in this temperature range [monoclinic space group $P2_1$ (1a in Fedorov notation) below T(lock-in) = 46K)] – nearly 1000x greater.

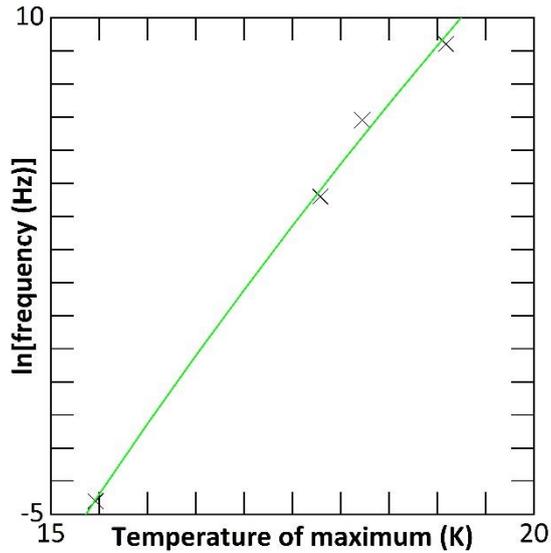

$$T_f = -14 \text{ K}; \; E_a = 0.33 \text{ eV}; \; f_0 = 1.04 \times 10^{+54} \text{ Hz}; \text{SSR} = 0.3688 \text{ Hz}^2$$

Figure 9: Our Vogel-Fulcher fit to the original data of Siegwarth et al. [82], note that attempt frequency $f_0$ is unrealistic.

To conclude, the reanalysis of literature suggests that this is a relaxor-type ferroelectric with low-temperature transition which for dc measurements is below absolute zero. This explanation will require experimental confirmation that maybe the subject of a future paper.

## IV. Conclusions

In summary, we present evidence for two new relaxor-type QCP within tetragonal tungsten bronze, $K_3Li_2Ta_5O_{15}$, and in lead pyrochlore, $Pb_2Nb_2O_7$, using literature as well as new experimental data for KLT. In each case, they are revealed to be glassy relaxors with freezing temperatures at or below absolute zero, whose dielectric permittivity peaks occur at temperatures that can, therefore, be fine-tuned through absolute zero via frequency. Thus they may be of interest as glassy QCP systems, adding to the list of QCP candidates that already includes perovskites and hexaferrites. However further investigations, especially structurally, are required for both as similar to other cryogenic ferroelectric such as $CdTiO_3$ [106,107] the controversy will continue.

## V. Acknowledgements

Facilities used for the RUS measurements were provided through grant no. EP/I036079/1 to MAC from the Engineering and Physical Sciences Research Council. Work at the University of St. Andrews supported by EPSRC grant EP/P024637/1. R.M.S. would like to thank the EPSRC for provision of a studentship via the doctoral training grant (EP/N509759/1).

We would like to thank A. M. Z. Slawin at the University of St Andrews for single crystal XRD on some other KLT samples.

References

[1] S. E. Rowley, L. J. Spalek, R. P. Smith, M. P. M. Dean, M. Itoh, J. F. Scott, G. G. Lonzarich, and S. S. Saxena, Nat. Phys. **10**, 367 (2014).

[2] J. F. Scott, A. Schilling, S. E. Rowley, and J. M. Gregg, Sci. Technol. Adv. Mater. **16**, 36001 (2015).

[3] S. E. Rowley, Y.-S. Chai, S.-P. Shen, Y. Sun, A. T. Jones, B. E. Watts, and J. F. Scott, Sci. Rep. **6**, 25724 (2016).

[4] S.-P. Shen, J.-C. Wu, J.-D. Song, X.-F. Sun, Y.-F. Yang, Y.-S. Chai, D.-S. Shang, S.-G. Wang, J. F. Scott, and Y. Sun, Nat. Commun. **7**, 10569 (2016).

[5] P. Chandra, G. G. Lonzarich, S. E. Rowley, and J. F. Scott, Reports Prog. Phys. **80**, 112502 (2017).

[6] H. B. Cao, Z. Y. Zhao, M. Lee, E. S. Choi, M. A. Mcguire, B. C. Sales, H. D. Zhou, J.-Q. Yan, and D. G. Mandrus, APL Mater. **3**, 62512 (2015).

[7] S. E. Rowley, M. Hadjimichael, M. N. Ali, Y. C. Durmaz, J. C. Lashley, R. J. Cava, and J. F. Scott, J. Phys. Condens. Matter **27**, 395901 (2015).

[8] J. C. Lashley, J. H. D. Munns, M. Echizen, M. N. Ali, S. E. Rowley, and J. F. Scott, Adv. Mater. **26**, 3860 (2014).

[9] F. Kagawa, N. Minami, S. Horiuchi, and Y. Tokura, Nat. Commun. **7**, 10675 (2016).

[10] F. Kagawa, S. Horiuchi, and Y. Tokura, Crystals **7**, 106 (2017).

[11] S. E. Rowley, T. Vojta, A. T. Jones, W. Guo, J. Oliveira, F. D. Morrison, N. Lindfield, E. Baggio Saitovitch, B. E. Watts, and J. F. Scott, Phys. Rev. B **96**, 020407(R) (2017).

[12] A. Rotaru, D. C. Arnold, A. Daoud-Aladine, and F. D. Morrison, Phys. Rev. B **83**, 184302 (2011).

[13] A. Rotaru and F. D. Morrison, J. Therm. Anal. Calorim. **120**, 1249 (2015).

[14] D. Viehland, S. J. Jang, L. E. Cross, and M. Wuttig, J. Appl. Phys. **68**, 2916 (1990).

[15] R. Pirc and R. Blinc, Phys. Rev. B **76**, 1 (2007).

[16] B. E. Vugmeister, Phys. Rev. B **73**, 174117 (2006).

[17] M. Trainer, Am. Assoc. Phys. Teach. **69**, 966 (2001).

[18] R. J. D. Tilley, Int. J. Refract. Met. Hard Mater. **13**, 93 (1995).

[19] A. Simon and J. Ravez, Comptes Rendus Chim. **9**, 1268 (2006).

[20] X. Zhu, M. Fu, M. C. Stennett, P. M. Vilarinho, I. Levin, C. A. Randall, J. Gardner, F. D. Morrison, and I. M. Reaney, Chem. Mater. **27**, 3250 (2015).

[21] R. Guo, A. S. Bhalla, C. A. Randall, Z. P. Chang, and L. E. Cross, J. Appl. Phys. **67**, 1453 (1990).

[22] J. F. Scott, A. Shawabkeh, W. F. Oliver, A. C. Larson, and P. J. Vergamini, Ferroelectrics **104**, 85 (1990).

[23] M.-S. Kim, J.-H. Lee, J.-J. Kim, H. Y. Lee, and S.-H. Cho, Integr. Ferroelectr. **69**, 11 (2005).

[24] G. Y. Kang and J. K. Yoon, J. Cryst. Growth **193**, 615 (1998).

[25] C.-Y. Jang, J.-H. Lee, J.-J. Kim, S.-H. Cho, and H. Y. Lee, J. Electroceramics **13**, 847 (2004).

[26] A. Peter, I. Hajdara, K. Lengyel, G. Dravecz, L. Kovacs, and A. Toth, J. Alloys Compd. **463**, 398 (2008).

[27] S. C. Abrahams, P. B. Jamieson, and J. L. Bernstein, J. Chem. Phys. **54**, 2355 (1971).

[28] C. Li, J. Han, J. Wang, L. Zhang, and H. Zhao, Chinese Phys. Lett. **14**, 468 (1997).

[29] Y. Xu, Z. Li, W. Li, H. Wang, and H. Chen, Phys. Rev. B **40**, 11902 (1989).

[30] Z. Lu, J.-P. Bonnet, J. Ravez, and P. Hagenmuller, J. Mater. Sci. **30**, 5819 (1995).

[31] W. Zhong, P. Zhang, H. Zhao, Z. Han, and H. Chen, Chinese Phys. Lett. **10**, 183 (1993).

[32] R. D. Shannon, Acta Crystallogr. **A32**, 751 (1976).

[33] T. Fukuda, Jpn. J. Appl. Phys. **9**, 599 (1970).

[34] M. Takashige, S.-I. Hamazaki, M. Tsukioka, F. Shimizu, H. Suzuki, and S. Sawada, J. Phys. Soc. Japan **62**, 1486 (1993).

[35] M. Takashige, S. Kojima, S.-I. Hamazaki, F. Shimizu, and M. Tsukioka, Jpn. J. Appl. Phys. **32**, 4384 (1993).

[36] M. Tsukioka, S. Hamazaki, M. Takashige, F. Shimizu, H. Suzuki, and S. Sawada, J. Phys. Soc. Japan **61**, 4669 (1992).

[37] S. Kojima, S. Hamazaki, M. Tsukioka, and M. Takashige, J. Phys. Soc. Japan **62**, 1097 (1993).

[38] S. Mori, N. Yamamoto, Y. Koyama, S. Hamazaki, and M. Takashige, Phys. Rev. B **52**, 9117 (1995).

[39] H. Ivett, A Kálium-Lítium-Niobát Kristály Tulajdonságai És Hibaszerkezete Tartalomjegyzék, PhD thesis, University of Pécs, 2013.

[40] H. M. Rietveld, J. Appl. Crystallogr. **2**, 65 (1969).

[41] B. H. Toby, J. Appl. Crystallogr. **34**, 210 (2001).

[42] A. C. Larson and R. B. Von Dreele, "General Structure Analysis System (GSAS)", Los Alamos National Laboratory Report LAUR 86-748 (1994).

[43] S. E. Rowley, M. Hadjimichael, M. N. Ali, Y. C. Durmaz, J. C. Lashley, R. J. Cava, and J. F. Scott, J. Phys. Condens. Matter **27**, (2015).

[44]   A. Migliori and J. L. Sarrao, in Reson. Ultrasound Spectrosc. Appl. to Physics, Mater. Meas. Nondestruct. Eval. (Wiley, New York, 1997), p. 201.

[45]   J. Schiemer, L. J. Spalek, S. S. Saxena, C. Panagopoulos, T. Katsufuji, A. Bussmann-Holder, J. Köhler, and M. A. Carpenter, Phys. Rev. B **93**, 54108 (2016).

[46]   D. M. Evans, J. A. Schiemer, M. Schmidt, H. Wilhelm, and M. A. Carpenter, Phys. Rev. B **95**, (2017).

[47]   M. A. Carpenter, E. K. H. Salje, and C. J. Howard, Phys. Rev. B **85**, 224430 (2012).

[48]   M. A. Carpenter and Z. Zhang, Geophys. J. Int. **186**, 279 (2011).

[49]   M. A. Carpenter, C. J. Howard, R. E. A. McKnight, A. Migliori, J. B. Betts, and V. R. Fanelli, Phys. Rev. B **82**, 134123 (2010).

[50]   M. Weller, G. Y. Li, J. X. Zhang, T. S. Kê, and J. Diehl, Acta Metall. **29**, 1047 (1981).

[51]   A. S. Nowick and B. S. Berry, (1972).

[52]   W. F. Oliver and J. F. Scott, Ferroelectrics **117**, 63 (1991).

[53]   P. W. Young and J. F. Scott, Phase Transitions **6**, 175 (1986).

[54]   J. F. Scott, S.-J. Sheih, and T. Chen, Ferroelectrics **117**, 21 (1991).

[55]   J. F. Scott and S.-J. Sheih, J. Phys. Condens. Matter **2**, 8553 (1990).

[56]   J. F. Scott, S. A. Hayward, and M. Miyake, J. Phys. Condens. Matter **17**, 5911 (2005).

[57]   C. Filipic, Z. Kutnjak, R. Lortz, A. Torres-Pardo, M. Dawber, and J. F. Scott, J. Phys. Condens. Matter **19**, 236206 (2007).

[58]   J. Gardner and F. D. Morrison, Dalt. Trans. **43**, 11687 (2014).

[59]   J. Gardner, F. Yu, C. Tang, W. Kockelmann, W. Zhou, and F. D. Morrison, Chem. Mater. **28**, 4616 (2016).

[60]   G. H. Olsen, U. Aschauer, N. A. Spaldin, S. M. Selbach, and T. Grande, Phys. Rev. B **93**, 180101(R) (2016).

[61]   T. Fukuda, Jpn. J. Appl. Phys. **8**, 122 (1969).

[62]   S.-L. Xu, J.-H. Lee, J.-J. Kim, H. Y. Lee, and S.-H. Cho, Mater. Sci. Eng. **B99**, 483 (2003).

[63]   R. Flores-Ramirez, A. Huanosta, E. Amano, R. Valenzuela, and A. R. West, Ferroelectrics **99**, 195 (1989).

[64]   E. Nakamura, T. Mitsui, and J. Furuichi, J. Phys. Soc. Japan **18**, 1477 (1963).

[65]   B. E. Jun, S. C. Song, D. J. Kim, C. S. Kim, G. S. Jeen, H. K. Kim, J. N. Kim, and Y. Hwang, J. Korean Phys. Soc. **42**, S1252 (2003).

[66]   J. S. Kim, D. J. Kim, and J. N. Kim, J. Korean Phys. Soc. **32**, S316 (1998).

[67]   D. C. Sinclair and A. R. West, Phys. Rev. B **39**, 13486 (1989).

[68]   V. Hornebecq, J. M. Reau, and J. Ravez, Solid State Ionics **127**, 231 (2000).

[69]    L. Benitez and J. M. Seminario, J. Electrochem. Soc. **164**, E3159 (2017).

[70]    Y. C. Chen, C. Y. Ouyang, L. J. Song, and Z. L. Sun, J. Phys. Chem. C **115**, 7044 (2011).

[71]    M. A. Carpenter and E. K. H. Salje, Eur. J. Mineral. **10**, 693 (1998).

[72]    W. Rehwald, Adv. Phys. **22**, 721 (1973).

[73]    M. A. Carpenter, J. F. J. Bryson, G. Catalan, S. J. Zhang, and N. J. Donnelly, J. Phys. Condens. Matter **24**, 45902 (2012).

[74]    R. E. A. McKnight, T. Moxon, A. Buckley, P. A. Taylor, T. W. Darling, and M. A. Carpenter, J. Phys. Condens. Matter **20**, 75229 (2008).

[75]    J. Herrero-Albillos, P. Marchment, E. K. H. Salje, M. A. Carpenter, and J. F. Scott, Phys. Rev. B **80**, 214112 (2009).

[76]    M. A. Carpenter, J. Phys. Condens. Matter **27**, 263201 (2015).

[77]    L. G. Van Uitert, J. J. Rubin, and W. A. Bonner, IEEE J. Quantum Electron. **4**, 622 (1968).

[78]    J. S. Abell, K. G. Barraclough, I. R. Harris, A. W. Vere, and B. Cockayne, J. Mater. Sci. **6**, 1084 (1971).

[79]    R. Ubic and I. M. Reaney, J. Am. Ceram. Soc. **85**, 2472 (2002).

[80]    J. K. Hulm, Phys. Rev. **92**, 504 (1953).

[81]    G. Shirane and R. Pepinsky, Phys. Rev. **92**, 504 (1953).

[82]    J. D. Siegwarth, W. N. Lawless, and A. J. Morrow, J. Appl. Phys. **47**, 3789 (1976).

[83]    F. Jona, G. Shirane, and R. Pepinsky, Phys. Rev. **98**, 903 (1955).

[84]    I. S. Zheludev, in *Phys. Cryst. Dielectr.* (Springer Heidelberg, 2012), p. 78.

[85]    W. N. Lawless, Phys. Rev. B **19**, 3755 (1979).

[86]    N. N. Kolpakova, P. Charnetzki, W. Nawrochik, P. P. Syrnikov, and A. O. Lebedev, J. Exp. Theor. Phys. **94**, 395 (2002).

[87]    A. Jayaraman, G. A. Kourouklis, A. S. Cooper, and G. P. Espinosa, J. Phys. Chem. **94**, 1091 (1990).

[88]    T. R. R. Shrout and S. L. L. Swartz, Mater. Res. Bull. **18**, 663 (1983).

[89]    W. N. Lawless, A. C. Anderson, and F. Walker, Ferroelectrics **37**, 627 (1981).

[90]    N. N. Kolpakova, B. Hilczer, and M. Wiesner, Phase Transitions **47**, 113 (1994).

[91]    N. N. Kolpakova and P. Czarnecki, J. Exp. Theor. Phys. **100**, 964 (2005).

[92]    N. N. Kolpakova, J. Exp. Theor. Phys. **96**, 538 (2003).

[93]    N. N. Kolpakova, P. Czarnecki, W. Nawrocik, M. P. Shcheglov, P. P. Syrnikov, and L. Szczepańska, Phys. Rev. B **72**, 24101 (2005).

[94]    E. Buixaderas, S. Kamba, J. Petzelt, M. Savinov, and N. N. Kolpakova, Eur. Phys. J. B **19**, 9 (2001).

[95]     Z. G. Ye, N. N. Kolpakova, J.-P. Rivera, and H. Schmid, Ferroelectrics **124**, 275 (1991).

[96]     N. N. Kolpakova, M. Wiesner, G. Kugel, and P. Bourson, Ferroelectrics **190**, 179 (1997).

[97]     N. N. Kolpakova, P. Czarnecki, W. Nawrocik, M. P. Shcheglov, and A. O. Lebedev, Ferroelectrics **291**, 125 (2003).

[98]     N. N. Kolpakova, R. Margraf, and M. Polomska, J. Phys. Condens. Matter **6**, 2787 (1994).

[99]     C. F. Clark, W. N. Lawless, and A. S. Bhalla, Jpn. J. Appl. Phys. **24**, 266 (1985).

[100]    X. W. Dong, K. F. Wang, S. J. Luo, J. G. Wan, and J. –M. Liu, J. Appl. Phys. **106**, 104101 (2009).

[101]    T. Katsufuji and H. Takagi, Phys. Rev. B **69**, 64422 (2004).

[102]    W. N. Lawless, Rev. Sci. Instrum. **42**, 561 (1971).

[103]    M. T. Sebastian, R. Ubic, and H. Jantunen, in *Microw. Mater. Appl.* (Wiley, Hoboken, NJ; Chichester, 2017), p. 181.

[104]    T. Sekiya, A. Tsuzuki, S. Kawakami, and Y. Torii, Mater. Res. Bull. **24**, 63 (1989).

[105]    K. Sreedhar and A. Mitra, J. Am. Ceram. Soc. **82**, 1070 (1999).

[106]    A. I. Lebedev, Phys. Solid State **51**, 802 (2009).

[107]    B. J. Kennedy, Q. Zhou, and M. Avdeev, J. Solid State Chem. **184**, 2987 (2011).